\documentclass[pre,twocolumn,showpacs,amsmath,amssymb,floatfix]{revtex4}

\usepackage{graphicx}
\usepackage{hyperref}
\usepackage{t1enc}

\newcommand{\dd}{{\text{d}}}

\newcommand{\vecq}{{\bf q}}
\newcommand{\vecp}{{\bf p}}
\newcommand{\vecr}{{\bf r}}

\begin{document}

\title{
  Tsallis thermostatistics for finite systems: a Hamiltonian approach
}

\author{Artur B. Adib}
  \email{artur@dartmouth.edu}
\affiliation{
  Department of Physics and Astronomy, Dartmouth College,
  Hanover, NH 03755, USA
}

\author{Andr\'{e} A. Moreira}
  \email{auto@fisica.ufc.br}
\author{Jos\'{e} S. Andrade Jr.}
  \email{soares@fisica.ufc.br}
\author{Murilo P. Almeida}
  \email{murilo@fisica.ufc.br}
\affiliation{
  Departamento de F\'{\i}sica, Universidade Federal do Cear\'{a},
Caixa Postal 6030, 60455-760, Fortaleza, Cear\'{a}, Brazil
}

\date{\today}

\begin{abstract}
We show that finite systems whose Hamiltonians obey a
generalized homogeneity relation rigorously follow the nonextensive
thermostatistics of Tsallis. In the thermodynamical limit, however,
our results indicate that the Boltzmann-Gibbs statistics is always
recovered, regardless of the type of potential among interacting
particles. This approach provides, moreover, a one-to-one correspondence
between the generalized entropy and the Hamiltonian structure of a
wide class of systems, revealing a possible origin for the intrinsic
nonlinear features present in the Tsallis formalism that lead naturally
to power-law behavior. Finally, we confirm these exact results through
extensive numerical simulations of the Fermi-Pasta-Ulam chain of
anharmonic oscillators.
\end{abstract}


\maketitle

\section{Introduction}

Few quantities in physics play such a singular role in their theories
as the entropy in the Boltzmann-Gibbs (BG) formulation of statistical
mechanics: it provides, in an elegant and simple way, the fundamental
link between the microscopic structure of a system and its macroscopic
behavior.  Almost two decades ago, Tsallis \cite{tsallis-first} 
proposed the following entropy expression as a nonextensive generalization 
of the Boltzmann-Gibbs (BG) formalism for statistical mechanics,
\begin{equation} \label{entropy}
S_q = k \frac{ 1 - \int \dd \vecr [f(\vecr)]^q }{q-1},
\end{equation}
where $k$ is a constant, $q$ a parameter and $f(\vecr)$ a probability
distribution over the phase space variables $\vecr$. Instead of the
usual BG exponential, upon maximization with a fixed average energy
constraint, the above entropy gives a
power-law distribution,
\begin{equation} \label{phasedist}
f(\vecr) \propto \left[ 1 - (q-1) \beta E(\vecr) \right]^{\frac{1}{q-1}},
\end{equation}
where $\beta$ is a temperature-like parameter and $E$ is the energy of
the system. In the limit $q \rightarrow 1$, the above expressions
reduce to the familiar BG forms of entropy, $S\!\!=\!\!-k \int \dd \vecr
f(\vecr) \ln \! f(\vecr)$, and exponential distribution $f(\vecr) \propto
\exp(-\beta E)$. The consequences of this generalization are manifold
and far-reaching (for a recent review see e.g. \cite{tsallis-review}),
but its most notorious one is certainly the fact that $S_q$, along with
other thermodynamical quantities, are {\em nonextensive} for $q \neq 1$.

Whether such formalism can be verified experimentally or elucidated
from any previously established theoretical framework are obvious questions
that arise naturally. The answer to the first one seems to be affirmative,
as there are currently numerous references to systems that are better
described within the generalized approach of Tsallis than with the
traditional BG formalism \cite{tsallis-review,tsallis-list},
the underlying argument being usually based on a best fit to either numerical
or experimental data by choosing an appropriate value for the parameter $q$.
This is, by far, the most frequent approach towards Tsallis' distribution
and is surely not conclusive. Its justification from first principles, although
already addressed at different levels in previous studies
\cite{plastinos,rajagopal,wilk}, is nevertheless controversial and still
represents an open research issue.
For example, it has been
recently suggested that systems in contact with finite heat baths
should follow the thermostatistics of Tsallis \cite{plastinos}. This
derivation, however, relies on a particular {\it ansatz} for the density
of states of the bath and lacks a more direct connection with the entropy,
as pointed out in Ref.~\cite{ramshaw}. Also, it has been shown in 
Ref.~\cite{vives} that the Tsallis statistics, together with the biased
averaging scheme, can be mapped into
the conventional Boltzmann-Gibbs statistics by a redefinition of 
variables that results from the scaling properties of the Tsallis entropy.

In a recent study \cite{murilo}, a derivation of the generalized
canonical distribution is presented from first principles statistical
mechanics. It is shown that the particular features of a macroscopic
subunit of the canonical system, namely, the heat bath, determines the
nonextensive signature of its thermostatistics. More precisely, it is
exactly demonstrated that if one specifies the heat
bath to satisfy the relation
\begin{equation} \label{q-eq}
\frac{\dd}{\dd E} \left( \frac{1}{\beta(E)} \right) = q-1, 
\end{equation}
where $\beta \propto 1/(qkT)$ is a temperature-like parameter and $T$
is the thermodynamic temperature, the form of the distribution
Eq.~(\ref{phasedist}) is recovered \cite{murilo}. Equation~(\ref{q-eq})
is essentially equivalent to Eq.~(\ref{phasedist}). However, it reveals
a direct connection between the finite aspect of the many-particle
system and the generalized $q$-statistics
\cite{foot}.
It is analogous to state that, if the condition
of an infinite heat bath capacity is violated, the resulting canonical
distribution can no longer be of the exponential type and therefore
should not follow the traditional BG thermostatistics. Inspired by
these results, we propose here a theoretical approach for the
thermostatistics of Tsallis that is entirely based on standard
methods of statistical mechanics. Subsequently, we will not only
recover the previous observation that an adequate physical setting for
the Tsallis formalism should be found in the physics of finite systems,
but also derive a novel and exact correspondence between the Hamiltonian
structure of a system and its closed-form $q$-distribution, supporting
our findings through a specific numerical experiment.

\section{Theoretical framework}

We start by considering, in a shell of constant energy, a system whose
Hamiltonian can be written as a sum of two parts, viz.
\begin{equation}
H(\vecr)=H_1(\vecr_1)+H_2(\vecr_2,\vecr_3,\ldots),
\end{equation}
where $\vecr=(\vecr_1,\vecr_2,\vecr_3,\ldots)=(r_1,\ldots,r_{2N})$,
with $\vecr_1=(r_1,\ldots,r_{n_1})$, $\vecr_2=(r_{n_1+1},\ldots,r_{n_1+n_2})$,
$\vecr_3=(r_{n_1+n_2+1},\ldots,r_{n_1+n_2+n_3})$ and so on.
The fact that Tsallis distribution is a power-law instead of
exponential strongly suggests us to look for scale-invariant forms of
Hamiltonians \cite{alemany}. Furthermore, since scale-invariant Hamiltonians
constitute a particular case of homogeneous functions \cite{landau},
our approach here is to show that, if $H_2$ satisfies a generalized
homogeneity relation of the type
\begin{equation} \label{homog-rel}
  \lambda H_2(\vecr_2,\vecr_3,\ldots)=H_2(\lambda^{1/{a_2}} 
  \vecr_2,\lambda^{1/{a_3}} \vecr_3, \ldots),
\end{equation}
where $\lambda,a_2, a_3, \ldots$ are non-null real constants, then 
the correct statistics for $H_1(\vecr_1)$ is the one proposed by
Tsallis. The foregoing derivation is based on a simple scaling
argument, but we shall draw parallels to Ref.~\cite{murilo} whenever
appropriate.

The structure function (density of states) for $H_2$ at the energy level
$E$ is given by
\begin{equation}
\Omega_2(E)=\int \delta(H_2(\vecr_2,\vecr_3,\ldots)-E)\dd V_2,
\end{equation}
where $\dd V_2 =\dd^{n_2}\vecr_2\dd^{n_3}\vecr_3\ldots$ is the volume
element in the subspace spanned by $(\vecr_2, \vecr_3, \dots)$. For
systems satisfying equation~(\ref{homog-rel}), this function can be
evaluated taking $\lambda>0$ and computing
\begin{eqnarray}
\Omega_2(\lambda E_0)&=&\int \delta(H_2(\vecr_2,\vecr_3,\ldots)-\lambda E_0)\dd
V_2\nonumber \\
&=&\frac{1}{\lambda}\int \delta(\lambda^{-1}H_2(\vecr_2,\vecr_3,\ldots)-E_0)\dd
V_2\nonumber \\
&=&\frac{1}{\lambda}\int \delta(H_2(\lambda^{-\frac{1}{a_2}}
\vecr_2,\lambda^{-\frac{1}{a_3}}
\vecr_3,\ldots )-E_0)\dd V_2\nonumber \\
&=&\frac{1}{\lambda}\int \delta(H_2(\vecr'_2,\vecr'_3,\ldots)-E_0)
\lambda^{\frac{n_2}{a_2} +
\frac{n_3}{a_3} + \ldots} \dd V'_2\nonumber \\
&=&\lambda^{1/(q-1)}\Omega_2(E_0),
\end{eqnarray}
where we define 
\begin{equation}
\frac{1}{q-1} \equiv  \sum_{i=2,3,\ldots}\!\!\! \frac{n_i}{a_i}
\, - 1,  \label{q-micro}
\end{equation}
and utilize the notation $\vecr'_2=\lambda^{-1/a_2}\vecr_2$,
$\vecr'_3=\lambda^{-1/a_3}\vecr_3$, etc., and $\dd V'_2=\dd^{n_2}
\vecr'_2\dd^{n_3}\vecr'_3\ldots$ Hence, if $\Omega_2$ is defined 
at a value $E_0$, it is also defined at every
$E=\lambda E_0$, with $\lambda>0$. We can then write
\begin{equation}
\Omega_2(E)=\left(\frac{E}{E_0}\right)^{\frac{1}{q-1}}\Omega_2(E_0),
\end{equation}
and express the canonical distribution law over the phase space of $H_1$ as 
\begin{eqnarray}
f(\vecr_1)&=&\frac{\Omega_2(H-H_1(\vecr_1))}{\Omega(H)} \nonumber \\
&=&\frac{\Omega_2(H)}{\Omega(H)} \left(1-\frac{H_1(\vecr_1)}
{H}\right)^{\frac{1}{q-1}},
\label{phasedist-new}
\end{eqnarray}
where $H$ is the total energy of the joint system composed by $H_1$
and $H_2$, and $\Omega(H)$ is its structure function,
\begin{equation}
\Omega(H)=\int \delta(H_1(\vecr_1)+H_2(\vecr_2,\vecr_3,\ldots)-H)dV_1dV_2,
\end{equation}
where $dV_1$ and $dV_2$ are the infinitesimal volume elements of the
phase spaces of $H_1$ and $H_2$, respectively. Comparing
Eq.~(\ref{phasedist-new}) with the distribution in the form of
Eq.~(\ref{phasedist}), we get the following relation between $q$,
$\beta$ and $H$:
\begin{equation}
\beta=\frac{1}{(q-1)H}.
\end{equation}
Notice that one could reach exactly the same result using the
methodology proposed in Ref.~\cite{murilo}, i.e. by evaluating
$\beta(E)=\Omega_2'(E)/\Omega_2(E)$ at $E=H$, calculating $q-1$
through Eq.~(\ref{q-eq}) and inserting these quantities back in
Eq.~(\ref{phasedist}).

As already mentioned, previous studies have shown that the
distribution of Tsallis Eq.~(\ref{phasedist}) is compatible with some
anomalous ``canonical'' configurations where the heat bath is finite
\cite{plastinos} or composes a peculiar type of extended phase-space
dynamics \cite{andrade}. In our approach, the observation of Tsallis
distribution simply reflects the finite size of a thermal environment
with the property (\ref{homog-rel}), the thermodynamical
limit corresponding to $q\rightarrow 1$ in Eq.~(\ref{q-micro}). We emphasize
that, although similar conclusions could be drawn from
Refs.~\cite{plastinos,andrade}, the theoretical framework introduced here
permits us to put forward a rigorous realization of the $q$-thermostatistics:
it stems from the {\it weak} coupling of a system to a ``heat bath'' whose
Hamiltonian is a homogeneous function of its coordinates, {\em the value of
$q$ being completely determined by its degree of homogeneity, Eq.
(\ref{q-micro})}.
This provides also a direct correspondence between the parameter $q$
and the Hamiltonian structure through geometrical elements of its
phase space, viz. the surfaces of constant energy
$H_2=E$.

As a specific application of the above results, we investigate the
form of the momenta distribution law for a classical $N$-body 
problem in $d$-dimensions. The Hamiltonian of such a system can be 
written as
\begin{eqnarray} \label{hamilt-nbody}
H(\vecp,\vecq) & = & \frac{1}{2}\sum_{i=1}^{N}\vecp_i^2 + V(\vecq_1,
\ldots,\vecq_N) \nonumber \\
& = & H_1(\vecp_1,\ldots, \vecp_N) + H_2(\vecq_1,\ldots,\vecq_N),
\end{eqnarray}
where we define $H_1 \equiv \frac{1}{2}\sum_{i=1}^N\vecp_i^2$,
$\vecp_i=(p_{i1},\ldots,p_{id})$ is the linear momentum vector of 
an arbitrary particle $i$ (hence the number of degrees of freedom 
of the system $1$ is $n_1=Nd$), and $H_2$ (the ``bath'') is due to 
a homogeneous potential $V(\vecq)$ of degree $\alpha$, i.e., $\lambda
V(\vecq)=V(\lambda^{1/\alpha} \vecq)$ with $\vecq=(\vecq_1,\ldots,
\vecq_N)$. At this point, we emphasize that  the distinction between
``system'' and ``bath'' is merely formal and does not necessarily
involve a physical boundary. It relies solely on the fact that we 
can decompose the total Hamiltonian in two parts \cite{khinchin}. 
By making the correspondences $\vecr_{1}=(\vecp_1,\ldots,\vecp_N)$,
$\vecr_{2}=(\vecq_1,\ldots,\vecq_N)$, $n_1=Nd$, $n_2=Nd$, $a_1=2$ 
and $a_2=\alpha$, the homogeneity relation (\ref{homog-rel}) is 
satisfied. From Eq. (\ref{phasedist-new}) it then follows that
\begin{equation} \label{phasedist-f}
f(\vecp_1,\ldots,\vecp_N) \propto \left[ H - \frac{1}{2}\sum_{i=1}^N
\vecp_i^2 \right]^{\frac{1}{q-1}},
\end{equation}
where the nonextensivity measure $q$ is given by
\begin{equation} \label{q-nbody}
\frac{1}{q-1} = \frac{Nd}{\alpha} -1 \,\,\,\,\,\Rightarrow \,\,\,\,\,
q = \frac{Nd}{Nd-\alpha}.
\end{equation}
It is often argued that the range of the forces should play a
fundamental role in deciding between the BG or Tsallis formalisms to
describe the thermostatistics of an $N$-body system \cite{tsallis-review}.
For example, the scaling properties of the one-dimensional Ising model
with long-range interaction has been investigated analytically
\cite{vollmayr} and numerically \cite{salazar-toral} in the context
of Tsallis thermostatistics, whereas in Ref. \cite{vollmayr2} a
rigorous approach was adopted to study the nonextensivity of a more
general class of long-range systems in the thermodynamic limit (see below).
Recall that, for a $d$-dimensional system, an interaction is said to
be long-ranged if $-d \leq \alpha \leq 0$. Within this regime, the
thermostatistics of Tsallis is expected to apply, while for $\alpha<-d$
the system should follow the standard BG behavior \cite{tsallis-review}.
This conjecture is not confirmed by the results of the problem at hand. Indeed,
Eq.~(\ref{phasedist-f}) is consistent with the generalized $q$-distribution
Eq.~(\ref{phasedist}) no matter what the value of $\alpha$ is, as long as
it is non-null and $N$ is finite. In the limit $N \to \infty$, however,
we always get $q \to 1$, with the value of $\alpha$ determining the
shape of the curve $q=q(N)$. If $\alpha>0$, $q$ approaches the value
$1$ from above, while for $\alpha<0$ the value of $q$ is always less
than $1$. Therefore, for (ergodic) classical systems with $N$ particles
interacting through a homogeneous potential,
the {\em equilibrium} distribution of momenta always goes to the Boltzmann
distribution, $f(\vecr_{1}) \propto \exp[-\beta H_{1}(\vecr_{1})]$,
when $N \to \infty$.  This observation should be confronted with the
recent results of Vollmayr-Lee and Luijten \cite{vollmayr2}, who investigated
the nonextensivity of long-range (therein ``nonintegrable'') systems
with algebraically decaying interactions through a rigorous Kac-potential
technique. Contrary to the trend established by
the practitioners of Tsallis' formalism, those authors argue that it is possible
to obtain the nonextensive scaling relations of Tsallis without
resorting to an a priori $q$-statistics, the Boltzmann-Gibbs
prescription ($q=1$) being sufficient for describing long-range systems of the
type above. Even though our findings embody partially the same message (we are
not yet concerned about scaling relations), there are some caveats that prevent
their results from being straightforwardly applicable to our problem:
neither a system-size regulator for the energy
nor a cutoff function is present in our treatment. This is immediately in contrast
with their observation that the ``bulk'' thermodynamics
strongly depends on the functional form of the regulator. Moreover, by not
addressing the distribution function explicitly at finite system sizes, that
work has very little in common with the most interesting part of our study,
which might in fact explain some observations of the $q$-distribution.
Notwithstanding these differences, we believe that an investigation of the
scaling properties of the system studied here would elucidate from a different
perspective the connection of Tsallis thermostatistics with nonextensivity
and is certainly a very welcome endeavor.

It is important to stress here that the essential feature determining
the canonical distribution is the geometry of the phase space region
that is effectively visited by the system. In a previous work by Latora
et. al. \cite{latora}, the dynamics of a classical system of $N$ spins
with infinitely long-range interaction is investigated through
numerical simulations, and the results indicate that if the
thermodynamic limit ($N\to \infty$) is taken before the infinite-time
limit ($t\to \infty$), the system does not relax to the
Boltzmann-Gibbs equilibrium. Instead, it displays anomalous behavior
characterized by stable non-Gaussian velocity distributions and
dynamical correlation in phase space. This might be due to the
appearance of metastable state regions that have a fractal nature with
low dimension. In our theoretical approach, however, it is assumed
that {\it the infinite-time limit is taken before the thermodynamic
limit}. As a consequence, metastable or quasi-stationary states like
the ones observed by Latora {\it et al.} \cite{latora} with a
particular long-range Hamiltonian system cannot be predicted within
the framework of our methodology. Whether this type of dynamical
behavior can be generally and adequately described in term of the
nonextensive thermostatistics of Tsallis still represents an open
question of great scientific interest.

\section{Numerical experiments}

In order to corroborate our method, we investigate
through numerical simulation the statistical properties of a linear
chain of anharmonic oscillators. Besides the kinetic term, the
Hamiltonian includes both on-site and nearest-neighbors quartic
potentials, i.e.
\begin{equation} \label{hamilt-osc}
H = \sum_{i=1}^{N}\frac{p_i^2}{2} + \sum_{i=1}^{N} \frac{q_i^4}{4} 
+ \sum_{i=1}^{N} \frac{\left( q_{i+1} - q_i \right)^4 }{4}.
\end{equation}
The choice of this system is inspired by the so-called
Fermi-Pasta-Ulam (FPU) problem, originally devised to test whether
statistical mechanics is capable or not to describe dynamical systems
with a small number of particles \cite{fpu}. From Eq.~(\ref{hamilt-osc}), 
we obtain the equations of motion and integrate them  numerically together
with the following set of initial conditions:
\begin{eqnarray} \label{init-cond}
q_i(0) & = & 0, \nonumber \\
p_i(0) & = & \sum_{k=1}^{4} \cos \left( \frac{2 \pi i k}{N} + \psi_k 
\right), ~~~ i=1,\ldots,N
\end{eqnarray}
where $\psi_k$ is a random number within $[0,2\pi)$. Undoubtedly, a
rigorous analysis concerning the ergodicity of this dynamical system
would be advisable before adopting the FPU chain as a plausible case
study. This represents a formidable task, even for such a simple
problem \cite{fpu-ford}. For our practical purposes, it suffices,
however, to test if the system displays equipartition among its linear
momentum degrees of freedom, since this is one of the main signatures
of ergodic systems. Indeed, one can show from the so-called Birkhoff-Khinchin
ergodic theorem that, for (almost) all trajectories \cite{berdichevsky},
\begin{equation} \label{gen-equip}
\lim_{t\rightarrow \infty} \left\langle p_i \frac{\partial H}{\partial p_i}
\right\rangle (t) =
\frac{\Gamma(E)}{\Omega(E)} \equiv \Theta, ~~~ i=1,\ldots,N
\end{equation}
where $\Gamma(E)=\int_{H(\vecr)<E}\dd V$ is the volume of the phase
space with $H<E$, $\langle A \rangle(t) = (1/t)\int_0^t \dd \tau A(\tau)$
denotes the the usual time average of an observable $A$, and $\Theta$ stands
for the absolute temperature of the {\em whole system}, $\Theta=4E/(3N)$
(cf. \cite{bannur}). We then follow the time evolution of the quantities
\begin{equation} \label{Q-test}
Q_i(t) \equiv \frac{1}{\Theta}
\left\langle p_i \frac{\partial H}{\partial p_i} \right\rangle (t),
~~~ i=1,\ldots,N
\end{equation}
to check if they approach unity as $t$ increases. From the results of
our simulations with different values of $N$ and several sets of
initial conditions, we observe in all cases the asymptotic behavior,
$Q_i(t) \to 1$ as $t \to \infty$. This procedure also indicates a good
estimate for the relaxation time of the system, $\tau \approx 10^5$,
so we shall consider our statistical data only for $t>10^5$, with a
typical observation time in the range $10^7 - 10^8$, after thermalization.

\begin{figure} 
\includegraphics[width=230pt]{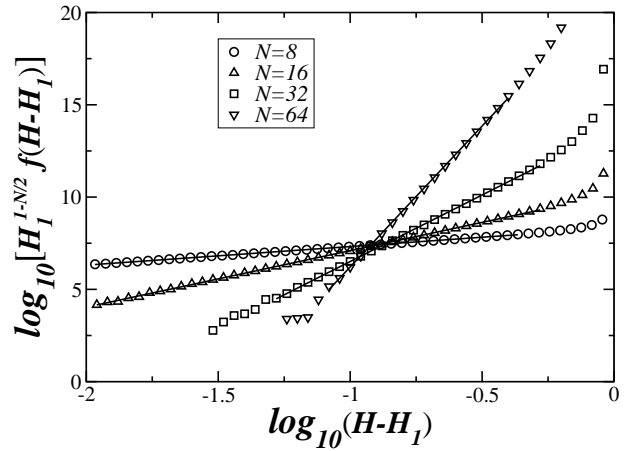}
\caption{
Logarithmic plot of the rescaled distribution $H_{1}^{1-N/2}f(H-H_{1})$
as a function of the transformed variable $H-H_{1}$ for $N=8$ (circles),
$16$ (triangles up), $32$ (squares), and $64$ (triangles down) anharmonic
oscillators. The solid straight lines are the best fit to the
simulation data of the expected power-law behavior Eq.~(\ref{phasedist-f}).
The slopes are 1.0068 (1.0), 3.07 (3.0), 7.21 (7.0), and 15.27 (15.0)
for $N=8$, $16$, $32$, and $64$, respectively, and the numbers in
parentheses indicate the expected values obtained from Eq.~(\ref{q-nbody}).
The departure from the power-law behavior at the extremes of the curves
is due to finite-time sampling.
}
\end{figure}
In Fig.~1 we show the logarithmic plot of the distribution
$f(\vecp_1,\vecp_2,\ldots)$ against the transformed variable $H-H_{1}$
for systems with $N=8$, $16$, $32$, and $64$ oscillators. Based on the
above result, we assume ergodicity and compute the distribution of
momenta from the fluctuations in time of $H_{1}$ through the relation
\begin{equation}
f(H_1) \propto H_1^{(N/2)-1} f(\vecp_1,\vecp_2,\ldots),
\end{equation}
where the $H_1^{(N/2)-1}$ factor accounts for the degeneracy of the
momenta consistent with the magnitude of $H_1$ (cf. \cite{khinchin}).
Indeed, we observe in all cases that the fluctuations in $H_{1}$ follow
very closely the prescribed power-law behavior Eq.~(\ref{phasedist-f}), 
with exponents given by Eq.~(\ref{q-nbody}). These results, therefore,
provide clear evidence for the validity of our dynamical approach to
the generalized thermostatistics.

\section{Conclusion} \label{conclusion}

In conclusion, we have shown that the generalized formalism of Tsallis
can be applied to homogeneous Hamiltonian systems to engender an
adequate theoretical framework for the statistical mechanics of finite
systems.
Of course, we do not expect that our approach can explain the whole
spectrum of problems in which Tsallis statistics can be applied.
However, our exact results clearly indicate
that, as far as homogeneous Hamiltonian systems are concerned,
the range of the interacting potential should play no role in the equilibrium
statistical properties of a system in the thermodynamic limit
\cite{inhomog}.
Under these conditions, the conventional BG thermostatistics remains
valid and general, i.e., for the specific class of homogeneous
Hamiltonians investigated here, the thermodynamic limit ($N\to
\infty$) leads always to BG distributions.

\acknowledgments

A. B. Adib thanks the Departamento de F\'{\i}sica at Universidade Federal
do Cear\'{a} for the kind hospitality during most part of this work and
Dartmouth College for the financial support. We also thank the Brazilian
agencies CNPq and FUNCAP for financial support.





\begin{thebibliography}{99}

\bibitem{tsallis-first}
C. Tsallis, J. Stat. Phys. {\bf 52}, 479 (1988).

\bibitem{tsallis-review}
C. Tsallis, ``Nonextensive statistical mechanics and thermodynamics:
Historical background and present status,'' in {\em Nonextensive
Statistical Mechanics and its Applications,} S. Abe and Y. Okamoto
(Eds.) (Springer-Verlag, Berlin, 2001); also in Braz. J. Phys.  {\bf
29}, 1 (1999).

\bibitem{tsallis-list}
See \url{http://tsallis.cat.cbpf.br/biblio.htm} for an updated bibliography.

\bibitem{plastinos}
A. R. Plastino and A. Plastino, Phys. Lett. A {\bf 193}, 140 (1994).

\bibitem{rajagopal}
S. Abe and A. K. Rajagopal, Phys. Lett. A {\bf 272}, 341 (2000);
J. Phys. A {\bf 33}, 8733 (2000).

\bibitem{wilk}
G. Wilk and Z. Wlodarczyk, Phys. Rev. Lett. {\bf 84}, 2770 (2000).

\bibitem{ramshaw}
J. D. Ramshaw, Phys. Lett. A {\bf 198}, 122 (1995).

\bibitem{vives} E. Vives and A. Planes, Phys. Rev. Lett. {\bf 88}, 020601
(2002).

\bibitem{murilo}
M. P. Almeida, Physica A {\bf 300}, 424 (2001).

\bibitem{foot} Observing that $\beta(E)\equiv \Omega'_2(E)/\Omega_2(E)$, where
$\Omega_2(E)$ is the structure function of the heat bath, and
$\Omega'_2(E)$ is its derivative, and integrating Eq.~(\ref{q-eq})
with the initial condition $\Omega(0)=0$ we get that $\Omega_2(E)=K E^{1/(q-1)}$,
where $K$ is a constant. This implies that the structure function is
a finite power of $E$ for $q \ne 1$, and therefore the phase space is
finite dimensional.

\bibitem{alemany}
It is worth mentioning that a related connection between
scale-invariant thermodynamics and Tsallis statistics was also
proposed by P. A. Alemany, Phys. Lett. A, {\bf 235} 452 (1997),
although the approach adopted by the author is not based on the more
fundamental ergodic arguments presented here.

\bibitem{landau}
L. D. Landau and E. M. Lifshitz, {\em Mechanics}, 3rd. Ed.  (Reed
Educ. and Prof. Pub., 1981). In this reference, ``scale invariance''
is under the name of ``mechanical similarity''.

\bibitem{andrade}
J. S. Andrade Jr., M. P. Almeida, A. A. Moreira and G. A. Farias,
Phys. Rev. E {\bf 65}, 036121 (2002).

\bibitem{khinchin}
A. I. Khinchin, {\em Mathematical Foundations of Statistical
Mechanics} (Dover, New York, 1949).

\bibitem{vollmayr} B. P. Vollmayr-Lee and E. Luijten,  Phys. Rev. Lett. {\bf
85}, 470 (2000).

\bibitem{salazar-toral} R. Salazar and R. Toral, Phys. Rev. Lett. {\bf 83},
4233 (1999).

\bibitem{vollmayr2} B. P. Vollmayr-Lee and E. Luijten,  Phys. Rev. E {\bf 63},
031108 (2001).

\bibitem{latora} V. Latora, A. Rapisarda and C. Tsallis, Phys. Rev. E {\bf 64},
056134 (2001).

\bibitem{fpu}
E. Fermi, J. Pasta, S. Ulam and M. Tsingou, {\em Studies of nonlinear
problems I}. Los Alamos preprint LA-1940 (7 November 1955); Reprinted
in E. Fermi, {\em Collected Papers}, Vol. II (Univ.  of Chicago Press,
Chicago, 1965) p. 978.

\bibitem{fpu-ford}
J. Ford, Phys. Rep. {\bf 213}, 271 (1992).

\bibitem{berdichevsky}
V. L. Berdichevsky, {\em Thermodynamics of Chaos and Order} (Addison
Wesley, 1997).

\bibitem{bannur}
V. M. Bannur, P. K. Kaw and J. C. Parikh, Phys. Rev. E {\bf 55}, 2525 (1997).

\bibitem{inhomog}
For recent investigations on the inhomogeneous case, see e.g.
C. Tsallis, Physica A {\bf 302}, 187 (2001), and V. Latora,
A. Rapisarda, and C. Tsallis, Phys. Rev. E {\bf 64}, 056134 (2001).

\end{thebibliography}
\end{document}